\documentstyle[12pt]{article}

\def\al{\nu}
\def\ba{\begin{array}}
\def\ea{\end{array}}
\def\bea{\begin{eqnarray}}
\def\eea{\end{eqnarray}}
\def\bea*{\begin{eqnarray*}}
\def\eea*{\end{eqnarray*}}
\def\be{\begin{equation}}
\def\ee{\end{equation}}

\begin{document}
\
\vspace{0.5cm}
\begin{flushright}
Preprint ITEP-TH-20/97\\
hep-th/9706045
\end{flushright}
\vspace{1cm}
\renewcommand{\thefootnote}{\arabic{footnote}}

\begin{center}
{ \Large \bf Duality in 2D Spin Models on Torus}
\end{center}
\vspace{1cm}
\bigskip
\begin{center} { Anatolij I. Bugrij \footnote
{e-mail: abugrij@gluk.apc.org }}
\end{center}
\begin{center}
{ Bogolyubov Institute for Theoretical Physics}
\end{center}
\begin{center}
 { 252143, Kiev, Ukraine}
\end{center}
\bigskip

\begin{center} { Vitalij N. Shadura}
\end{center}
\begin{center}
{ Institute of Theoretical and Experimental Physics}
\end{center}
\begin{center}
 { 117259, Moscow, Russia}
\end{center}
\begin{center}
{ and}
\end{center}
\begin{center}
{ Bogolyubov Institute for Theoretical Physics}
\end{center}
\begin{center}
 { 252143, Kiev, Ukraine}
\end{center}
\bigskip
\bigskip

\begin{abstract}
\begin{sloppypar}

Method  of derivation of the duality relations for two-dimensional
$Z(N)$-symmetric spin models on finite square lattice wrapped
on the torus  is proposed. As example, exact  duality relations
for the  nonhomogeneous Ising model ($N=2$) and
\hfill \break the $Z(N)$-Berezinsky-Villain model
are obtained.


 \end{sloppypar}
\end{abstract}
\thispagestyle{empty}

\newpage
\pagenumbering{arabic}

\section{Introduction}

Study of  duality properties in  statistical mechanics and
quantum field theory models is important method for non-perturbative
investigation of their phase diagram and field content.
Duality transformation was discovered by Kramers and Wannier [1]
in the two-dimensional Ising model.
Kadanoff and Ceva [2]  generalized   the   Kramers-Wannier duality relation
 to the nonhomogeneous case (the coupling constants are
arbitrary functions of lattice site coordinates)
with spherical boundary conditions:
\be \bigl(\prod_{\widetilde{r},\mu} \sinh 2
\widetilde{K}_\mu
(\widetilde{r})\bigr)^{-1/4}\widetilde{Z}\bigl[
\widetilde{K}\bigr]
=
\bigl(\prod_{r,\mu} \sinh 2 K_\mu(r)\bigr)^{-1/4}Z\bigl[K\bigr],
\quad \mu=x,y,
\label {di1}
\ee
\be
\sinh 2K_x(r)\cdot\sinh2\widetilde{K}_{-y}(\widetilde{r})=1,
\quad
\sinh 2K_y(r)\cdot\sinh2\widetilde{K}_{-x}(\widetilde{r})=1.
\label {di2}
\ee
We denote site coordinates, functions and functionals  on
the dual lattice by "tilda" :
$ \widetilde{r}, \,$
$\widetilde{\sigma}(\widetilde{r}), \,$ $ \widetilde{K}_\mu(\widetilde{r}),
\, $ $\widetilde{H}[\widetilde{K},\widetilde{\sigma}], \,$
$\widetilde{Z}[\widetilde{K}], \,\, \dots\, $ .
A site coordinate on the dual lattice coincides with a coordinate
of the plaquet center on  the original lattice:
$\widetilde{r}=r+(\widehat{x}+\widehat{y})/2$
and   coupling constants
 $\widetilde{K}_{-\nu}(\widetilde{r})=
\widetilde{K}_{\nu}(\widetilde{r}-\widehat \nu)$
($\widehat \nu=\widehat{x}$,  $\widehat{y}$ are  the unit vectors along
the horisontal $X$ and vertical $Y$ axes).

As was already mentioned in [1,2],  relation (\ref{di1})
can not be understood  literally.
So, for example, using the method of comparing  high- and
low-temperature expansions for deriving  duality relation (\ref{di1})
in the case of
 the periodical boundary conditions, it is hard to take into account
 and to compare the graphs
wrapping up the torus. In fact (\ref{di1})
is correct in the thermodynamic limit (for the specific free energy).  However
for the nonhomogeneous case the procedure of thermodynamic limit is rather
ambiguous.  In [2] this duality relation  was obtained for spherical
(nonphysical for the lattice) boundary conditions.

In [3], using global Bianchi identities for link formulation of the lattice
spin systems on the hypertorus, the contributions of the link variables on
the topological nontrivial loops on the hypertorus was selected in the
partition function, but the duality relations  was not formulated in obvious
form in this case.

 Since  duality is a popular  method of  non-perturbative investigation in
quantum field theory and statistical mechanics (for review see [10]),  it
is important to formulate a duality transformation for finite systems.
Recently, we have suggested [4,5]  exact duality relations for the
nonhomogeneous Ising model on a finite square lattice of size $n\times m$
wrapped on the torus:
\be
\prod_{\widetilde{r},\mu}\bigl(\sinh2\widetilde{K}_\mu(\widetilde{r})
\bigr)^{-1/4}{\widetilde Z}^{(\widetilde p_x,\widetilde p_y)}[\widetilde{K}]=
{1\over2}
\prod_{r,\mu}\bigl(\sinh2K_\mu (r)\bigr)^{-1/4}
\sum_{p_x,p_y=0}^{1}T^{\widetilde p_x,\widetilde p_y}_{p_x,p_y}
Z^{(p_x,p_y)}[K], \label {bi1}
\ee
Here  $Z^{(p_x,p_y)}[K]$ are partition
functions of the Ising model with corresponding combinations of the
periodical ($p_x,p_y=0$) and antiperiodical ($p_x,p_y=1$) boundary conditions
along  the horizontal $X$ and
vertical $Y$ axes:
\be Z^{(p_x,p_y)}[K]=\sum_{[\sigma]}\exp\bigl(
\sum_{r,\al}K_\al(r)\sigma(r)\nabla^{(p_\al)}_\al
\sigma(r)\bigr),
\label {bi2}
\ee
and
\be
\widehat{T}=\left(\ba{rrrr} \ 1&\ 1&\ 1&\ 1\\ \ 1&\ 1&-1&-1\\ \
 1&-1&\ 1&-1\\ \ 1&-1&-1&\ 1\ea\right),
\label {bi3}
\ee
where $r=(x,y)$    denotes the site coordinates on the square lattice of
size $n\times m$,  $x=1,\dots,n$ $y=1,\dots,m$;  $\sigma(r)=\pm1$;
$K_x(r)$ and $K_y(r)$ are  the coupling constants
along  corresponding
 axes.
The one-step shift operators  $\nabla_x$, $\nabla_y$
 act on $\sigma(r)$  in the following way
\be
\nabla_x\sigma(r)=\sigma(r+\widehat{x}),\quad\nabla_y\sigma(r)=\sigma
(r+\widehat{y}).
\label {sh1}
\ee
For   the periodical (antiperiodical) boundary conditions along $X$ and $Y$ axes
we have
\be
\nabla_x^{(p_x)} \sigma(n,y)=(-)^{p_x}\sigma(1,y),\quad
\nabla_y^{(p_y)} \sigma(x,m)=(-)^{p_y}\sigma(x,1).
\label {bi4}
 \ee

In Ref. [4] the duality relation (4) was proved
for  homogeneous  and
  weakly nonhomogeneous distributions of
the coupling constants.  We also have checked   the duality relation (4) for
lattices of small sizes by direct calculation on the computer.
As a corollary of  (4),   we
obtained [4,5] the duality relations  for the two-point correlation function
on the torus, for the partition functions of  the 2D Ising model  with
magnetic fields applied to the boundaries and   the 2D Ising model with
free, fixed and mixed boundary conditions.

In this paper we formulate  method of derivation  of the duality relations
for two-dimensional $Z(N)$-symmetric spin models on finite square lattice
wrapped on the torus.
As example, the duality relations for
the vector Potts model (the $N=2$ case is considered in detail) and
the $Z(N)$-Berezinsky-Villain model [6,7]
are obtained.
Without taking account of boundary conditions
duality relations for these models was obtained in [8,9] (for review see
[10]).  In principle suggested method it is not hard to generalize for
lattices with larger dimensions compactified on the hypertorus and the
lattice models with continious global or gauge symmetries.

To formulate the method let us introduce  definition of magnetic
dislocations connected with boundary conditions, "topological" charge of
dislocation
and gauge transformations of coupling constant configurations for
the vector Potts model.
The hamiltonian of this model one can write in the following form:
\be
-\beta{H}^{(p,q)}_{V}[K,\sigma]= {1\over 2}\sum_{r,\al}
\bigl(K_\al(r)\sigma^{*}(r)\nabla^{(p_\al)}_\al\sigma(r)+\mbox{c.c.}\bigr)
\label {v1}
\ee
where a spin variable takes $N$ values:
$\sigma(r)=\exp(i{{2\pi}\over N}l(r)),\quad $ $l(r)=0,...,\,N-1 $,
$\al= x,y$ and
 $p_x=p$,  $p_y=q$  ($p,q = 0,\,..., N-1$)
designate the cyclic boundary conditions for
 one-step shift operators
(\ref {sh1}):
\be
\nabla_x^{(p)} \sigma(n,y)=e^{i{{2\pi}\over
N}p}\sigma(1,y),\quad \nabla_y^{(q)} \sigma(x,m)=e^{i{{2\pi}\over
N}q}\sigma(x,1).
\label {b1}
\ee
These conditions have the following form for variable
$l(r)$:
 \be
l(n+1,y)=l(1,y)+p,\quad l(x,m+1)=l(x,1)+q.
\label {b2}
\ee
For the periodical boundary conditions we have
 $p=0$ and $q=0$.

Using (\ref{b1}),
Hamiltonian ${H}^{(p,q)}[K,\sigma]$  one can write as
Hamiltonian
${H}_D^{(0,0)}[K,d,\sigma]$ with the magnetic dislocation
${D}^{(p,q)}$ corresponding boundary conditions
 ${(p,q)}$ and with periodical boundary conditions for
spin variable
$\sigma(r)$:
$$
-\beta{H}^{(p,q)}[K,\sigma]=
-\beta
{H}_D^{(0,0)}[K,d,\sigma]=
{1\over 2}\sum_{r,\nu}
\bigl[K_\nu(r)\exp\bigl(i{{2\pi}\over N}d^{(p,q)}_\nu(r)\bigr)\,
\sigma^{*}(r)\nabla^{(0)}_\nu\sigma(r)+\mbox{c.c.}\bigr]
$$
\be
=\sum_{r,\al} K_\nu(r)\cos{{2\pi}\over N} (\Delta_\nu l(r)+ d^{(p,q)}_\nu(r)),
\label {v2}
\ee
Here  $\Delta_\nu=\nabla^{(0)}_\nu-1$ is difference derivative
with the periodical boundary conditions, vector fields
$K_\al(r)$
and  $d^{(p,q)}_\al(r)$, determined on the lattice bonds,
it is convenient to consider as the module and the phase
of the complex coupling constant.
The magnetic dislocation    ${D}^{(p,q)}$  is determined by the phase
\be
d^{(p,q)}_\nu(r) = (d_x(r),d_y(r))= (p\delta_{B_X}(r),q\delta_{B_Y}(r)\,),
\label {d1}
\ee
which is unequal zero along the boundary cycle
$B_X$  and $B_Y$, setting the space configuration of
the dislocation on the torus:
\be \delta_{B_X}(r)=
\sum_{r^\prime\in B_{X}} \delta^2(r-r^\prime),\quad \delta_{B_Y}(r)=
\sum_{r^\prime\in B_Y} \delta^2(r-r^\prime),
\label {d2}
\ee
where  $\delta^2(r-r^\prime)$ is Kronecker $\delta$-function and
$$
B_X= \left\{ (x,m),
x=1,...,n \right\}, \quad B_Y= \left\{ (n,y), y=1,...,m \right\}.
$$
The phase $d^{(p,q)}_\al(r)$ one can consider as density of a "topological"
charge
$Q_\nu $ of the magnetic dislocation.
This charge, for example, for dislocation
${D}^{(p,q)}$ is equal
\be Q_\nu= \sum_{r} d^{(p,q)}_\nu(r) =
(pn,qm).
\label {d3}
\ee
We will call  magnetic dislocations
${D}^{(p,q)}$  ($p,q =
0,\,..., N-1$) as basic magnetic dislocations.
Note that periodical boundary conditions ($p=q=0$) along all
cycles of the torus correspond the absence of the magnetic dislocations.
Nevertheless, for convenience we have introduce denotion
${D}^{(0,0)}$ for this case.

Hamiltonian
(\ref{v1}) has invariance relative to
$Z_N$-gauge transformations [11]
\be
\sigma^\prime(r)=e^{i{{2\pi}\over N}\phi(r)}\sigma(r),\quad
K^\prime_\mu(r)=e^{i{{2\pi}\over N}\phi(r)}K_\mu(r)e^{i{{2\pi}\over
N}\phi(r+\hat\mu)}, \label {G1}
\ee
where $\phi(r)$ has the periodical boundary conditions.
This invariance gives the following relation for partition function:
$$ Z_V^{(p,q)}[K]= \sum_{[\sigma]}
e^{-\beta{H}^{(p,q)}[K,\sigma]}=
\sum_{[\sigma^\prime]}e^{-\beta{H}^{(p,q)}[K^\prime,\sigma^\prime]}=
Z_V^{(p,q)}[K^\prime].
$$
Note that the gauge transformation of
 $l(r)$  and $d_\mu(r)$  in Hamiltonian
 (\ref{v2}) has form:
\be
l^\prime(r)=l(r)+\phi (r),\quad  d^{\prime}_\mu(r)=d^{(p,q)}_\mu(r)+\Delta_\mu\phi (r).
\label {G2}
\ee
These transformations lead to both the deformation of the basic magnetic
dislocations and the appearance of new closed dislocations.
Then
 $d^{\prime}_\mu(r)$ describes the field of  closed magnetic dislocations
on the torus.
It is obvious that the topological charge does not change at the gauge
transformation. For example, for Hamiltonian
${H}_D^{(0,0)}[K,d,\sigma]$ with dislocation ${D}^{(p,q)}$ we have
$$
Q^{\prime}_\mu=
\sum_{r} d^{(g)}_\mu(r) =\sum_{r} d^{(p,q)}_\mu(r)+ \sum_{r}
\Delta_\mu\phi (r).
$$
Here the periodical boundary conditions for
$\phi (r)$ lead to vanishing of the second term and $Q^{\prime}_\mu=Q_\mu$.
From here it follows that  the set of
coupling constant configurations
$\{[K,d^{(g)}]\}$ (contained  closed dislocations) one can
divide on the gauge-nonequivalent classes
$\Omega^{(p,q)}$  with corresponding value of
topological charge
$Q_\mu= (pn,qm)$.
Elements of class
$\Omega^{(p,q)}$ one can generate with help of gauge transformations
(\ref{G1}) from the basic magnetic dislocation
${D}^{(p,q)}$.

Let us briefly formulate the idea of suggested method.
Using the Fourier transformation method for derivation of the duality
relations, we obtain the expression with $\delta$-functions.  The solution of
the corresponding system of equations defines the relation between the
initial and dual spin variable.  Usually, for example, see [10], omitting the
problem of taking account of boundary conditions, the only one solution of
this system of equations is written.  However for the lattice model on the
torus we can find many solutions  of this system. These solutions one can
classify over the gauge-nonequivalent classes $\tilde \Omega^{(\tilde
p,\tilde q)}$ of coupling constant configurations for the dual model and also
 each class has  definite value of topological charge $\tilde Q_\mu=
(\tilde pn,\tilde qm)$.
Therefore  at dual transformation of the partition function
it is necessary to sum over all
the gauge-nonequivalent classes on the dual lattice
with coefficients, depending from boundary
conditions on the initial lattice.

\section {Vector Potts model}

Now, using  the method discussed in previous section, we derive
the duality relation for
 the vector Potts model.
Partition function (\ref{v2}) of this model one can
represented in the following form [9,10]:
\be
Z_V^{(p,q)}[K,d] =
\sum_{[\,l\,]} \exp\left\{-\beta{H}^{(p,q)}[K,l]\right\}=
\sum_{[\,l\,]}
\exp\left\{\sum_{r,\mu} K_\mu(r)\cos{{2\pi}\over N}
\bigl(\Delta_\mu l(r)+ d^{(p,q)}_\mu(r)\bigr)  \right\}=
\label {vd1}
\ee
\be
\sum_{[\,l\,]} \sum_{[\,t\,]} \exp(-\beta \tilde H [t]) \exp\left\{
i{{2\pi}\over N}\sum_{r,\mu}
t_\mu(r) \bigl(\Delta_\mu l(r)+ d^{(p,q)}_\mu(r)\bigr) \right\}=
\label {vd2}
\ee
\be
\sum_{[\,t\,]} \exp\bigl(-\beta \tilde H [t]+
i{{2\pi}\over N}\sum_{r,\mu}t_\mu(r)  d^{(p,q)}_\mu(r)\bigr)
\prod_r N\delta_N(\Delta_\mu t_\mu(r-\widehat\mu)),
\label {vd3}
\ee
where
$$
\sum_{[\,l\,]}=\prod_{r}(\sum_{l(r)=0}^{N-1}),\quad
\sum_{[\,t\,]}=\prod_{r}(\sum_{t_{\mu}(r)=0}^{N-1}).
$$
In (\ref{vd2}) we have made  the Fourier transformation to
vector field
$t_\mu(r)$
($t_\mu(r)=0,1,\,...,N-1 $).  $-\beta \tilde H[t]$
is Fourier-transform of Hamiltonian (11):
\be -\beta \tilde H
[t]=\sum_{k=0}^M\sum_{r,\mu}g^{(k)}_\mu(K)\cos^{k} {{2\pi}\over N}t_\mu(r).
\label {vd4}
\ee
Here
 $M=N/2$, if $N$ is even and
   $M=(N-1)/2$, if  $N$ is odd.
In (\ref{vd3}) $\delta_N(s)$ is Kronecker $\delta_N$-function
$\mbox{mod N}$: it is one if
 $s=NL$, where
 $L$ is integer and zero in another case.

 In order to get rid of
$\delta_N$-function in (\ref{vd3}) it is necessary to solve
equation
\be
\Delta_\mu t_\mu(r-\widehat\mu)=0\quad\mbox{/. mod N}.
\label {vd5}
\ee
Nontrivial solutions of this equation on the torus one can write
in the form:
\be
t^{(\alpha)}_\mu(r)=\epsilon_{\mu\nu} \Delta_\nu\tilde l(\tilde r-\widehat\nu)+
\epsilon_{\mu\nu}\tilde d^{(\alpha)}_\nu(\tilde r-\widehat\nu),
\label {vd6}
\ee
where index $\alpha$ numerates the solutions,
$\tilde l(\tilde r)=
0,1,\,...,N-1$ is defined on a site of the dual lattice  and
$\tilde d^{(\alpha)}_\nu(\tilde r)$ is
density of the topological charge (corresponding given solution
 $\alpha$)
of the field of the closed magnetic dislocations on the dual
lattice
\be
\tilde d^{(\alpha)}_\mu(\tilde r)
=\sum_{i\in Z_\alpha}  s^{(\alpha)}_i  \sum_{r^\prime\in
\Gamma_i}\epsilon_{\mu\nu} a_\nu(r^\prime)\delta^2(r-r^\prime),\quad
s^{(\alpha)}_i=0,1,...,N-1.
\label {vd7}
\ee
Here by analogy with
(\ref{d1}), (\ref{d2})
$\tilde d^{(\alpha)}_\mu(\tilde r)$
is defined on  bonds of the dual lattice.
For convenience we have written the dislocations on the dual lattice
by means of closed paths
$ \Gamma_i$  on the original lattice.
 $Z_\alpha$ in (\ref{vd7}) denotes subset of the paths
(corresponding to solution
 $\alpha$)  from set
$\Gamma$  of all closed paths on original lattice
  ($ \Gamma_i\in \Gamma$).
 Vector $a_\mu(r)=
e_\mu(r)$ if  the direction of circuit over path
 $\Gamma_i$ in site $r$
(the direction of circuit is counterclockwise)
coincides with  direction
of  the unit vector
$e_\mu(r)=\widehat{\mu}$ in this site,
otherwise $a_\mu(r)=-e_\mu(r)$.

Expression
 (\ref{vd7}) it is not hard to obtain, observing, that
solution
(\ref{vd6}) satisfies  by equation
(\ref{vd5}) on site $\tilde r$
when
$$
\epsilon_{\mu\nu} \Delta_\mu \tilde d^{(\alpha)}_\nu(\tilde
r-\widehat\mu-\widehat\nu)=0.
$$
This equation becomes the identity if m the following
 conditions are fulfilled:
$$ \tilde d^{(\alpha)}_y(\tilde r-\widehat y)= \tilde
d^{(\alpha)}_y(\tilde r-\widehat x-\widehat y),\quad \tilde
   d^{(\alpha)}_x(\tilde r-\widehat x)= \tilde d^{(\alpha)}_x(\tilde
r-\widehat x-\widehat y);
$$
$$
\tilde d^{(\alpha)}_y(\tilde r-\widehat y)=
\tilde d^{(\alpha)}_x(\tilde r-\widehat x ), \quad
\tilde d^{(\alpha)}_y(\tilde r-\widehat x-\widehat y)=
\tilde d^{(\alpha)}_x(\tilde r-\widehat x-\widehat y);
$$
$$
\tilde d^{(\alpha)}_y(\tilde r-\widehat y)=
-\tilde d^{(\alpha)}_x(\tilde r-\widehat x-\widehat y),\quad
\tilde d^{(\alpha)}_x(\tilde r-\widehat x )=
\tilde d^{(\alpha)}_y(\tilde r-\widehat x-\widehat y).
$$
Consistency of these solutions on some set of sites requires
that these sites belong to closed paths
$ \Gamma_i$ on the torus, that is these solutions must be "glued"
 in order to form the closed magnetic dislocatuons.

Let us denote by
$[\tilde d^{(\alpha)}]$  coupling constant configurations
on the dual lattice corresponding to the solution
(\ref{vd7}).
Depending on the number of the solution these configurations
cointain both the closed dislocations non-enveloping of the cycles of the
torus and the dislocations enveloping of ones.  The dislocations of the first
type remove by means of gauge transformations (\ref{G2}) on the dual lattice
and the dislocations of the second type can be trasformed to the basic
magnetic dislocations $\tilde {D}^{(\widetilde p,\widetilde q)}$.  This means that
all configurations $[\tilde d^{(\alpha)}] $ can be classified in the
gauge-nonequivalent classes $\tilde \Omega^{(\widetilde p,\widetilde q)}$ with
topological charge
$$ \tilde Q_\al= \sum_{\tilde r} \tilde d^{(\widetilde
p,\widetilde q)}_\al(\widetilde r) = (\tilde p n,\tilde q m),
$$ where $\widetilde p,
\widetilde q = 0,1,\,...,N-1$.

Since the duality relation connects the partition functions,
 which are the gauge-invariant quantities, at removal
$\delta_N$-functions in  (\ref {vd3})
we must sum over the gauge-nonequivalent solutions of equation
(\ref{vd5}):
\be
t^{(\widetilde p,\widetilde q)}_\mu(r)=
\epsilon_{\mu\nu} \Delta_\nu\widetilde l(\tilde r-\widehat\nu)+
\epsilon_{\mu\nu}\tilde d^{\,(\widetilde p,\widetilde q)}_\nu(\tilde r-\widehat\nu),
\label {vd8}
\ee
where $\tilde d^{\,(\widetilde p,\widetilde q)}_\mu$
is defined on the dual lattice by relations similar to
(\ref{d1})-(\ref{d3}).
Substituting these solutions  in
(\ref{vd3}),  we obtain
$$
Z^{(p,q)}_V[K,d] ={1\over
N}\sum_{\widetilde p,\widetilde q}\, \sum_{[\,\tilde l\,]} \exp\bigl(-\beta \tilde H
[ \Delta_\mu\tilde l+\tilde d^{\,(\tilde p,\tilde q)}_\mu]\bigr)
$$
$$
\exp\{i{{2\pi}\over N}\sum_{r,\mu} \epsilon_{\mu\nu} d^{(p,q)}_\mu(r)\bigl[
\Delta_\nu\tilde l(\tilde r-\widehat\nu)+ \tilde d^{\,(\widetilde p,\widetilde
q)}_\nu(\tilde r-\widehat\nu) \bigr] \}.
$$
Here we have introduced factor
 $1/N$ as
taking into account relation  (\ref{vd8}), it is not hard to note that
the sum over configurations
 $[l]$ in $N$ times more than the sum over $[t]$.
Remarking, that
$$
\sum_{r,\mu}\epsilon_{\mu\nu} d^{(p,q)}_\mu(r) \Delta_\nu\tilde l(\tilde
r-\widehat\nu)=0,
$$
   relation (\ref{vd8})
one can write in compact form
$$ Z^{(p,q)}_V[K,d] = {1\over N}\sum_{\widetilde p,\widetilde q}\,
\exp\bigl(i{{2\pi}\over N}\sum_{r,\mu}
\epsilon_{\mu\nu} d^{(p,q)}_\mu(r)
\tilde d^{\,(\widetilde p,\widetilde q)}_\nu(\tilde r-\widehat\nu)\bigr)
\tilde Z^{(\widetilde p,\widetilde q)}_V[\tilde K,\tilde d]=
$$
\be
{1\over N}\sum_{\tilde p,\tilde q}\,
\exp\bigl({i{{2\pi}\over N}(p
\widetilde q- q\widetilde p)}\bigr)
\tilde Z^{(\widetilde p,\widetilde q)}_V[\tilde K,\tilde d],
\label {vd9}
\ee
where
$$
\tilde Z^{(\widetilde p,\widetilde q)}_V[\tilde K,\tilde d]=
\sum_{[\,\tilde l\,]} \exp\bigl(-\beta \tilde H^{(\widetilde p,\widetilde q)}
[\tilde l,\tilde d ]\bigr)
$$
is the partition function of the model on the dual lattice.

 Let us in detail consider the case $N=2$.
Here Hamiltonian (\ref{v1})  coincides with
Hamiltonian (\ref{bi2}) of Ising model.
In this case from   (\ref{vd4})  one gets
$$
-\beta \tilde H_2^{(\widetilde p,\widetilde q)} [\tilde l ]= \sum_{r,\al}
\bigl[g^{(0)}_\al(\tilde K)+g^{(1)}_\al(\tilde K)\cos{{\pi}}\bigl(
\Delta^{\tilde p_\al} _\al\tilde
l(r)\bigr) \bigr].
$$
In order to find  coefficients
 $g^{(i)}_\mu(\tilde K)$  we use the inverse Fourier transformation
\be
\exp\bigl(\sum_k^M
g^{(k)}_\mu(K)\cos^{k}{{2\pi}\over N}\tilde t_\mu\bigr) ={1\over
N}\sum_{n=0}^{N-1} \exp\bigl(K_\nu\cos{{2\pi}\over N}n -i{{2\pi}\over N}n\tilde
t_\nu\bigr).
\label {vd10}
\ee
Here $\mu\neq\nu$.
Hence it is not hard to obtain for $N=2$:
$$
e^{2g^{(0)}_\mu(\tilde r)}={1\over2} \sinh 2K_\nu(r),\quad
e^{-2g^{(1)}_\mu(\tilde r)}= \tanh K_\nu(r)=e^{-2\tilde K_\mu(\tilde r)},
$$
where the last relation coinsides with
(\ref{di2}).
Using these relations, duality relation
(\ref{vd9}) one can represented in the form
\be
\prod_{r,\mu}\bigl(\sinh2K_\mu{(r)}\bigr)^{-1/4}Z^{(p,q)}[K]={1\over 2}
\prod_{\tilde r,\mu}\bigl(\sinh2\tilde K_\mu{(\tilde r)}\bigr)^{-1/4}
\sum_{\tilde p,\tilde q=0}^1 e^{i\pi(p\tilde q-q\tilde p)}
\tilde Z^{(\tilde p,\tilde q)}[\tilde K],
\label {vd11}
\ee
where
$$
\tilde Z^{(\widetilde p,\widetilde q)}[\tilde K]=\sum_{[ l ]}
\exp\sum_{\tilde r,\mu}
\bigl(\tilde K_\mu(\tilde r)\cos{{\pi}}\bigl(
\Delta^{\tilde p_\mu} _\mu\tilde l(r)\bigr) =
\sum_{[\sigma]}
\exp\sum_{\tilde r,\mu}
\bigl(\tilde K_\mu(\tilde r)\tilde \sigma(\tilde r)\nabla^{(\tilde p_\mu)}_\mu
\tilde \sigma(\tilde r)
\bigr),
$$
and $\tilde \sigma(\tilde r)=\pm 1$.
It is easy to verify that
(\ref{vd11}) coinsides with duality relation
(\ref{bi1}), since   the matrix
\be T_{\widetilde p_x,\widetilde p_y}^{p_x,p_y}=T^{p,q}_{\widetilde
p,\widetilde q}= e^{i\pi(p\widetilde q-q\widetilde p)}.
\label {bi5} \ee

\section{ The $Z(N)$-Berezinsky-Villain model}

Now let us consider the duality relation
for
the  $Z(N)$-symmetric Berezinsky-Villain model.
Partition function of this model one can write in the following form
[6,7]:
\be
Z_{BV}^{(p,q)}[K] = \sum_{[l]} e^{-\beta{H}_G^{(p,q)}[K,l]}=
\sum_{[l]}\sum_{[k]}\prod_{r,\mu}
\exp\left\{-{1\over 2}K_\mu(r)\bigl[{{2\pi}\over N} \Delta_\mu
l(r)-{2\pi}k_\mu (r) \bigl]^2\right\}, \label {g1}
\ee
where
$$
\sum_{[l]}=\prod_{r}\bigr(\sum_{l(r)=0}^{N-1}\bigl),\quad
\sum_{[k]}=
\prod_{r,\mu}\bigr(\sum_{k_\mu(r)=-\infty}^{\infty}\bigl).
$$
Here
$l(r)=0,...,\,N-1 $ is  on a site of the square lattice,
index
$(p,q)$ defines the boundary conditions
(\ref{b2})
ans the sum over
$k_\mu$
guarantes the periodicity of the Hamiltonian relative to shifts
$l\rightarrow  l(r)+ NL(r)$, where $L$ is integer.
By analogy with the partition function
of the vector Potts model
(\ref{g1}) one can rewrite  in term of the basic magnetic dislocations
 ${D}^{(p,q)}$:
\be
Z_{BV}^{(p,q)}[K,d]=
\sum_{[l]}\sum_{[k]}\prod_{r,\mu}
\exp\left\{-{1\over 2}K_\mu(r)\bigl[{{2\pi}\over N} (\Delta_\mu
l(r)+d^{(p,q)}_\mu(r))-{2\pi}k_\mu (r) \bigl]^2\right\},
\label {g2}
\ee
where $l(r)$  satisfies the periodical boundary conditions
and the density $d_\mu(r)$ of the topological charge
is determined by relations
(\ref{d1})-(\ref{d3}).

For derivation of the duality relation
let us  make  the  following transformations with (\ref{g2}):
 \be
 Z_{BV}^{(p,q)}[K,d]=
\sum_{[l]}\sum_{[k]}\prod_{r,\mu}
\exp\left\{-{1\over 2}K_\mu(r)\bigl[{{2\pi}\over N} (\Delta_\mu
l(r)+d_\mu(r))-{2\pi}k_\mu (r) \bigr]^2\right\}=
\label {gd1}
\ee
\be
(\prod_{r,\mu} N)^{1\over 2}
\sum_{[s]}\sum_{[k]}
\int D\theta \prod_{r,\mu}
\exp\left\{-{1\over 2}K_\mu(r)\bigl[
 (\Delta_\mu \theta (r)+{{2\pi}\over N}d_\mu(r))-2 \pi  k_\mu (r) \bigr]^2
+i{N\over 2} s(r){\theta(r)}\right\}=
\label {gd2}
\ee
\be
(\prod_{r,\mu}{N\over
{{2\pi
K_\mu(r)}}})^{{1\over 2}}
\sum_{[s]}\sum_{[t]}
\int D\theta \prod_{r,\mu}
 \exp\left\{-{1\over{2K_\mu(r)}}t^2_\mu(r)+i
 t_\mu(r)(\Delta_\mu \theta(r)+{{2\pi}\over N}d_\mu(r))
\right.
\label {gd3}
\ee
$$
\left.+i{N\over 2} s(r){\theta(r)}\right\}=
(\prod_{r,\mu}{N\over{{2\pi K_\mu(r)}}})^{1\over 2}
\sum_{[s]}\sum_{[t]}
 \prod_{r,\mu}
 \exp\left\{-{1\over{2K_\mu(r)}}t^2_\mu(r)+i{{2\pi}\over N}
 t_\mu(r)d_\mu(r)\right\}
$$
\be
\prod_{r}\delta \bigl(
 \sum_{\mu}\Delta_\mu
 t_\mu(r-\widehat \mu)-N s(r) \bigr),
\label {gd4}
 \ee
where
$$
\sum_{[t]}=
\prod_{r,\mu}\bigr(\sum_{t_\mu(r)=-\infty}^{\infty}\bigl),\quad
   \int D\theta=\prod_{r}\bigr(\int_{0}^{2\pi}{d\theta(r)\over {2\pi}}\bigl).
$$
For derivation (\ref{gd2})-(\ref{gd4})
we have used  the summation  formula
 $$ {{2\pi}\over
 N}\sum_{l=0}^{N-1}\delta(\theta-{{2\pi}\over N}l0)=
 \sum_{s=-\infty}^{\infty} e^{i N s\theta}, \quad 0\leq\theta\leq 2\pi,
 $$
 the identity
 $$
 \sum_{k=-\infty}^{\infty}
 \exp\bigl[-{1\over 2}K\bigl(f-2\pi k \bigr)^2\bigr]= {1\over{\sqrt{2\pi K}}}
 \sum_{t=-\infty}^{\infty}
 \exp\bigl(-{1\over {2K}}t^2 +itf \bigr)
 $$
and the definition of
 $\delta$-function
 $$
 \int_{0}^{2\pi}{d\theta(r)\over {2\pi}}e^{i\theta l}= \delta(l).
 $$

In order to  take off the
$\delta$-function in (\ref{gd4})
it is necessary to solve the equation
  \be
 \sum_{\mu}\Delta_\mu
 t_\mu(r-\widehat \mu)=N s(r).
\label {gd5}
 \ee
Analysis of the solutions of this equation is similar to analysis of
(\ref{vd5})-(\ref{vd8}) and leads to the followig expressions for
the gauge-nonequivalent solutions
\be
t^{(\widetilde p,\widetilde q)}_\mu(r)=
\epsilon_{\mu\nu} \Delta_\nu\tilde l(\tilde r-\widehat\nu)+
\epsilon_{\mu\nu}\tilde d^{\,(\widetilde p,\widetilde q)}_\nu(\tilde r-\widehat\nu)
-N\epsilon_{\mu\nu}\tilde k_\nu(\tilde r-\widehat\nu).
\label {gd6}
\ee
\be
s(r)=
\epsilon_{\mu\nu} \Delta_\mu
 \tilde k_\nu(\tilde r-\widehat\nu-\widehat\mu).
\label {gd7}
\ee
These solutions are the basic magnetic dislocations
 $\tilde {D}^{(\widetilde p,\widetilde q)}$
on the dual lattice  in corresponding
the gauge-nonequivalent classes
$\tilde \Omega^{(\widetilde p,\widetilde q)}$  with the topological
charge
 $Q_\mu= (\tilde p n,\tilde q m)$
$(\tilde p, \tilde q = 0,1,\,...,N-1)$.
Then,  taking off $\delta$-functions in (\ref{gd4}),
it is necessary to sum over all these solutions.
In result we obtain  the duality relation for
the $Z(N)$-Berezinsky-Villain model
\be
\bigl(\prod_{r,\mu}{
{2\pi K_\mu(r)}\over N}\bigr)^{1\over4}
Z_{BV}^{(p,q)}[K,d]=
{1\over N}
 \sum_{\tilde p,\tilde q}\,
\exp\bigl({i{{2\pi}\over N}(p
\tilde q- q\tilde p)}\bigr)
\bigl(\prod_{\widetilde r,\mu}{{2\pi \widetilde K_\mu
 (\widetilde r)}\over N}\bigr)^{1\over4}
\tilde Z^{(\widetilde p,\widetilde q)}_{BV}[\widetilde K,\widetilde  d],
\label {gd8}
 \ee
where
$$
K_\mu(r) \widetilde K_{-\nu} (\widetilde r)=\bigr({N\over{2\pi}}\bigl)^2,
\quad \mu\neq\nu.
$$

As it is shown in [8] this model at
$N=2$  corresponds to the Ising model, what  is consistent
with  our result
(\ref{gd8}), which coincides with
(3) in this case.
\medskip
V.S. thanks Dr. A.~Morozov for the hospitality and the exellent conditions
at ITEP, where this paper has been finished.

\bigskip

\centerline{\bf References}

\bigskip

\begin{enumerate}
\item H.A.~Kramers, G.H.~Wannier, Phys. Rev.
{\bf 60}, 252 (1941).
\item L.P.~Kadanoff, H.~Ceva, Phys. Rev. B {\bf 3}, 3918
(1971).
\item G.G.~Batrouni, M.B.~Halpern,
Phys. Rev. D{\bf 30}, 1775 (1984).
\item A.I.~Bugrij, V.N.~Shadura,
Zh. Eksp. Teor. Fiz.
 {\bf 109}, 1024 (1996).
\item A.I.~Bugrij, V.N.~Shadura,
Pis`ma Zh. Eksp. Teor. Fiz.
 {\bf 63}, 369 (1996).
\item V.L.~Berezinsky, Ph.D. Dissertation, Landau ITP, 1971.
\item J.~Villain, J.Phys. C{\bf 36}, 581 (1975).
\item Al.B.~Zamolodchikov,
Zh. Eksp. Teor. Fiz.
 {\bf 75}, 341 (1978).
 \item V.S.~Dotsenko,
Zh. Eksp. Teor. Fiz.
 {\bf 75}, 1083 (1978).
\item R.~Savit, Rev. Mod. Phys. {\bf 52}, 453 (1980).
 {\bf 63}, 369 (1996).
\item E.~Fradkin, V.A.~Huberman, S.H.~Shenker, Phys. Rev.
 B {\bf 18}, 4789 (1978).
\end{enumerate}

 \end{document}